\documentclass[twocolumn,showpacs,preprintnumbers,amsmath,amssymb,superscriptaddress]{revtex4}

\usepackage{graphicx}
\usepackage{dcolumn}
\usepackage{bm}

\begin{document}


\title{Surface dark solitons in nonlocal nonlinear media}

\author{XingHui Gao}
\affiliation{Laboratory of Photonic Information Technology, South
China Normal University, Guangzhou 510631, P. R. China}
\author{Luohong Zhou}
\affiliation{Laboratory of Photonic Information Technology, South
China Normal University, Guangzhou 510631, P. R. China}
\author{ZhenJun Yang}
\affiliation{Laboratory of Photonic Information Technology, South
China Normal University, Guangzhou 510631, P. R. China}
\author{Xuekai Ma}
\affiliation{Laboratory of Photonic Information Technology, South
China Normal University, Guangzhou 510631, P. R. China}
\author{Daquan Lu}
\affiliation{Laboratory of Photonic Information Technology, South
China Normal University, Guangzhou 510631, P. R. China}
\author{Wei Hu} \email[Corresponding author:
]{huwei@scnu.edu.cn}
\affiliation{Laboratory of Photonic Information Technology, South
China Normal University, Guangzhou 510631, P. R. China}

\date{\today}

\begin{abstract}
We predict the existence of surface dark solitons at the interface between a self-defocusing
nonlocal nonlinear medium and a linear medium. The fundamental and higher-order surface dark solitons can exist when the linear refractive index of the self-defocusing media is much larger than that of the linear media. The fundamental solitons are stable and the stabilities of higher-order solitons depend on both nonlocality degree and propagation constant.
\end{abstract}

.

\maketitle

Optical surface waves, a special type of waves that is localized
at the interface between two different media, are widely used in sensing physical, chemical and
biological agents to explore intrinsic and extrinsic properties of
material surface. In the presence of nonlinearity, some kinds of surface solitons
have been found theoretically and experimentally, such as in local Kerr media \cite{Tomlinson-ol-1980,Mihalachea-pio-1989},
waveguide arrays \cite{Makris-ol-2005,Katashov-prl-2006},  photorefractive media \cite{Quirino-pra-1995,Cronin-Golomb-ol-1995}, and metamaterials \cite{Lazarides-pre-2008}. Surface dark solitons
existing at the interface between the local nonlinear media and the linear media have been studied\cite{Skinner1991-josab,Chen1992-pra}.

Nonlinear surface solitons can exist not only at
the interface in local nonlinear media but also in nonlocal nonlinear media. Varies types of surface
solitons in nonlocal nonlinear media have been found and studied, such as bright
surface fundamental solitons\cite{Alfassi-2007-PRL}, incoherent surface
solitons \cite{Alfassi-2009-PRA}, surface dipoles
\cite{Ye-2008-PRA,V.Kartashov-2009-OL} and surface vortices
\cite{Ye-2008-PRA}.   In self-defocusing nonlocal media, the bright surface solitons and ring surface solitons have been predicted and studied\cite{Kartashov2007-ol,Kartashov2007-OE}.  However, can surface dark solitons exist in self-defocusing
nonlocal media? There is no study till now, though dark solitons in nonlocal self-defocusing medium were studied theoretically\cite{Nikolov2004-OL,Yaroslav2007-ol,Ouyang2009-oe} and experimentally\cite{Dreischuh2006-PRL,Fischer2006-ol,Ghofraniha2007-PRL,Conti2009-PRL}.

In this Letter,we address the existence of surface dark solitons (SDSs) at the
interface between a self-defocusing thermal medium and a linear medium. We find that a SDS in a self-defocusing nonlocal media is identical with the half part of a dark soliton in the same bulk media when the linear refractive index of the self-defocusing media is much larger than that of the linear media. The similar relation between bright surface solitons and bulk solitons in nonlocal nonlinear media is firstly given in Ref.\cite{Yang2011-OE} under the same assumption.

We start our analysis by considering a TE polarized laser beam with an complex envelope $q$
propagating along the $z$ axis in the vicinity of the interface formed by a self-defocusing
thermal nonlinear medium and a linear medium. The propagation of laser waves is described by the dimensionless (1+1)-D nonlocal nonlinear Schr\"{o}dinger equation: \\
(i)in self-defocusing nonlinear media, i.e. $x\leq0$
\begin{subequations}\label{eq1}
\begin{eqnarray}
 i\frac{\partial q}{\partial z}+\frac{1}{2}\frac{\partial^{2}q}{\partial x^{2}}+\Delta n q &=&0,  \\
       \Delta n  - w^{2}_{m}\frac{\partial^{2}\Delta n}{\partial x^{2}} &=& - |q|^{2},
\end{eqnarray}
\end{subequations}
(ii)in linear media, i.e. $x>0$
\begin{equation}\label{eq2}
    i\frac{\partial q}{\partial z}+\frac{1}{2}\frac{\partial^{2}q}{\partial x^{2}}-qn_{d}=0.
\end{equation}
Here the transverse $x$ and longitudinal $z$ coordinates are scaled in terms of beam width and diffraction length, respectively.
$\Delta n$ denotes the nonlinear perturbation of refractive
index in the self-defocusing thermal medium. $n_{d}>0$  describes the difference of unperturbed refractive index between the self-defocusing thermal medium and
the linear medium. The parameter $w_{m}$ stands for the characteristic length of the nonlocal material response. We
use $w_{m}/w_{0}$ represents the degree of the nonlocality, where $w_{0}$ is the beam width.

For the TE polarized wave, the continuity condition for transverse fields is  $q(x=+0)=q(x=-0)$.  Following the experimental instance \cite{Alfassi-2007-PRL}, the continuity condition for the nonlinear refractive index is assumed as $\partial\Delta n/\partial x|_{x=0}=0$, which can be achieved when the interface between nonlinear and linear media is thermally insulating.  We assume $n_d>>1$, which is easily satisfied in an actual physical system\cite{Yang2011-OE}. When $n_d>>1$ almost all the fields of laser beams will locates in the higher-index media\cite{Alfassi-2007-PRL,Alfassi-2009-PRA,Ye-2008-PRA,V.Kartashov-2009-OL,Kartashov2007-ol,Kartashov2007-OE,Yang2011-OE}, so the boundary condition in the lower-index linear medium is $q(x\rightarrow+\infty)=0$.

\begin{figure}[htb]
\includegraphics[width=8cm]{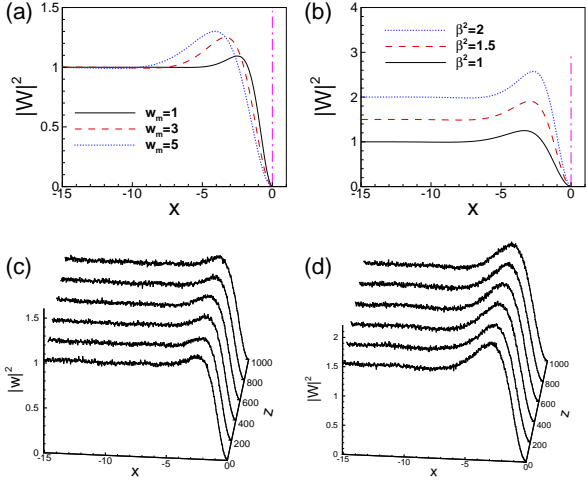}
\caption{(Color online)Profiles of fundamental SDSs for (a)  $w_m=1,3, 5$, $\beta^2=1$;(b)$w_m=3$, $\beta^2=1, 1.5, 2$.  Propagation of fundamental SDSs with (c)$w_{m}=1$, $\beta^2=1$;(d)$w_{m}=3$, $\beta^2=1.5$. Purple dash-dotted lines show the interface.}
\label{funadamental}
\end{figure}

The asymptotic behaviors of SDSs when $x\rightarrow -\infty$ are similar to that of bulk dark solitons\cite{Ouyang2009-oe}. If the stationary solution is in the form $q(x,z)=W(x)e^{i\beta z}$, we have $W(x\rightarrow-\infty)=\sqrt{-\beta}$. Therefore the propagation constant $\beta$ must be negative. Consequently,  one also have $\Delta n(x\rightarrow-\infty)=\beta$. In an actual thermal nonlinear system, $\Delta n$ is proportional to the change of temperature due to the absorption of the wave energy. For a uniform distribution of optical intensity $I=|q|^2=|W|^2$(i.e. in the  vicinity of $x\rightarrow-\infty$),  the distribution of temperature could not be uniform because the temperature gradient is necessary for the  transfer of heat energy. As a result, the exact uniform background for surface and bulk dark solitons in thermal nonlinear media can not be achieved. A much broad beam is used to approximate the uniform background (same in local case) in experiments\cite{Dreischuh2006-PRL,Fischer2006-ol}.  For the theocratical study of SDSs in this Letter, $\partial \Delta n /\partial x|_{x\rightarrow-\infty}=0$ and $\partial q /\partial x|_{x\rightarrow-\infty}=0$ are used in numerical simulations for the convenience.

\begin{figure}[htb]
\includegraphics[width=7.5cm]{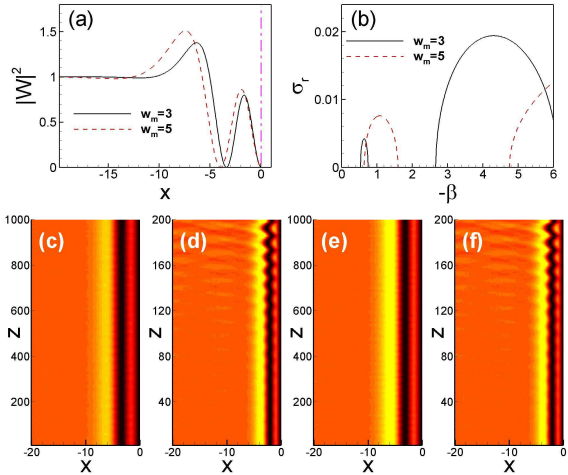}
\caption{(Color online)(a)Profiles of second-order SDS with $w_m=3,5$, $\beta^2=1$.(b)The perturbation growth
rate versus  propagation constant for $w_{m}=3, 5$. Propagation of second order SDS for (c)$w_{m}=3$, $\beta^2=1$; (d)$w_{m}=3$, $\beta^2=4.5$; (e) $w_{m}=5$, $\beta^2=2.1$; (f) $w_{m}=5$, $\beta^2=1$.}\label{second}
\end{figure}

\begin{figure}[htb]
\includegraphics[width=7.5cm]{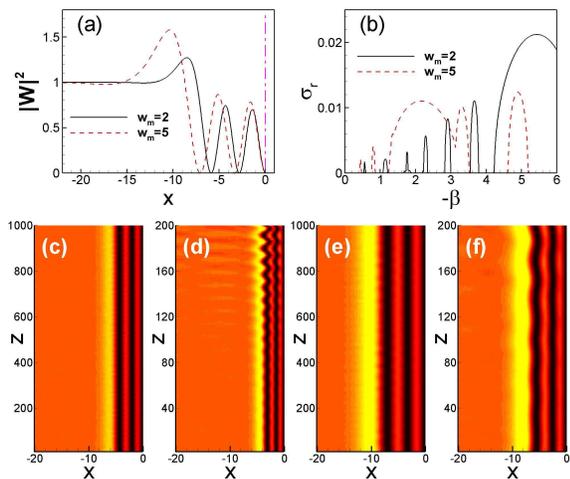}
\caption{(Color online)(a)Profiles of third-order SDS with $w_{m}=2, 5$, $\beta^2=1$. (b)The perturbation growth
rate versus propagation constant for $w_{m}=2, 5$. Propagation of third-order SDS
 for (c) $w_{m}=2$, $\beta^2=2$;(d) $w_{m}=2$, $\beta^2=4.6$; (e) $w_{m}=5$, $\beta^2=1$; (f) $w_{m}=5$, $\beta^2=2$. }.\label{third}
\end{figure}

\begin{figure}[htb]
\includegraphics[width=7.5cm]{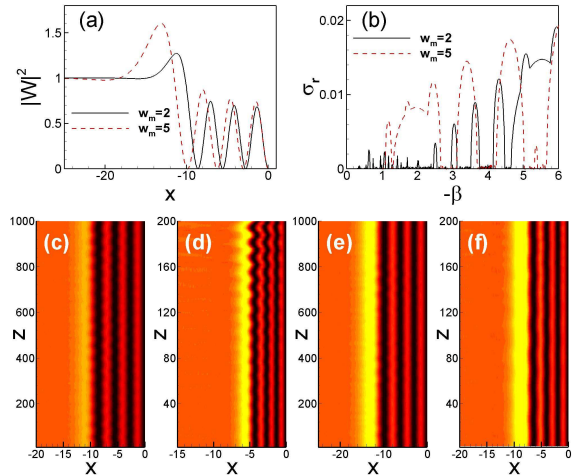}
\caption{(Color online)(a)Profiles of forth-order SDSs for $w_{m}=2,5$, $\beta^2=1$. (b)The perturbation growth
rate versus propagation constant for $w_{m}=2, 5$.  Propagation of the forth-order SDSs
for (c)$w_{m}=2$, $\beta^2=1$; (d) $w_{m}=2$, $\beta^2=5$; (e) $w_{m}=5$, $\beta^2=1$; (f) $w_{m}=5$, $\beta^2=3.4$. }\label{forth}
\end{figure}

We search for stationary solutions for  Eqs.(\ref{eq1}) and (\ref{eq2}) numerically in the form $q(x,z)=W(x)e^{i\beta z}$, where $W(x)$ is a real function and the propagation constance $\beta<0$.  The profiles of fundamental and higher-order SDSs are shown in Figs.\ref{funadamental}-\ref{forth}. We find the fields of SDSs resides all in the self-defocusing media with higher refraction index. The profiles of SDSs are proved to be identical with the half part of corresponding  bulk dark solitons, which are obtained numerically from Eqs.(\ref{eq1}) for a bulk medium($-\infty <x <\infty$).

To elucidate the linear stability of SDSs, we search for perturbed solutions
in the form $q(x,z)=[W(x)+a(x,z)]e^{i\beta z}$ and $\Delta n=N(x)+b(x,z)$,  where $|a(x,z)|\ll |W(x)|$ and $|b(x,z)|\ll |N(x)|$ are real perturbations. Substituting the
perturbed solutions into Eqs.(\ref{eq1}) and (\ref{eq2}),  one can get the linearized equations around stationary solutions  $W(x)$ and $N(x)$\\
(i) in noninear media, $x<0$,
\begin{subequations}
\begin{eqnarray}
 i\frac{\partial a}{\partial z}+ \frac{1}{2}\frac{\partial^{2}a}{\partial x^{2}}-\beta a -(N a+U b)=0 \\
 b-w_{m}^{2}\frac{\partial^{2}b}{\partial x^{2}}= U^{*}a+U a^{*}
\end{eqnarray}
\end{subequations}
(ii) in linear medis, $x>0$
\begin{equation}
i\frac{\partial a}{\partial z}+ \frac{1}{2}\frac{\partial^{2}a}{\partial x^{2}}-(\beta+n_d) a=0.
\end{equation}
The perturbations $a(x,z)$ and $b(x,z)$ can grow with a complex rate $\sigma$ upon propagation. Following the method in Ref.\cite{Armaroli2009-pra}, one can get a linear eigenvalue problem for $\sigma$. The eigenvalue problem  has been solved numerically and   the instability growth rate $\sigma_r$ (real part of $\sigma$) are shown in Figs.\ref{second} -\ref{forth}. In addition, to confirm the results of linear stability analysis, we perform numerical simulations base on Eqs.(\ref{eq1}) and (\ref{eq2}). The solutions of SDSs obtained numerical are used as the incident profiles with noise in the form $q(x,z=0)=W(x)[1+\rho(x)]$, where $\rho(x)$ is a random function with a Gaussian distribution and variance $\sigma_{noise}^{2}=0.01$.

Figure \ref{funadamental} shows results for the fundamental SDSs with different nonlocality degrees and different propagation constants. As the nonlocality degree increases, the shoulder of SDSs increases and moves away from the interface.  The width of the intensity valley increases too. With increasing of $\beta$, the intensity of dark background increase and the width of valley decreases.  The linear stability analysis shows the fundamental SDSs are always stable. The numerical simulations with noise-added incident profiles, shown in Figs.\ref{funadamental}(c) and (d), confirm the stability od SDSs.

Figure \ref{second} shows results for the second-order SDSs with different nonlocality degrees. There are two valleys in the second-order SDSs as shown in Fig.\ref{second}(a).  As the nonlocality degree increases, the shoulder of the second-order SDSs increases and moves away from the interface.  The width of two valleys increases together.  The results of the linear stability analysis are shown in Fig.\ref{second}(b). One can see that the stability of the second-order SDSs depends on both $w_m$ and $\beta$.  The numerical simulations with noise-added incident profiles are shown in Figs.\ref{second}(c) - (f).  In Figs.\ref{second}(c) and (e), the propagation are stable over 1000 times diffraction length, as predicated by the stability analysis[see Fig.\ref{second}(b)]. In Figs.\ref{second}(d) and (f) which are predicated unstable, the waves begin to oscillate over 100 times diffraction length..

Figure \ref{third} shows results for the third-order SDSs with different nonlocality degrees. There are three valleys in the third-order SDSs as shown in Fig.\ref{third}(a).  As the nonlocality degree increases, the shoulder of the third-order SDSs increases and moves away from the interface.  The width of three valleys increases together.  The results of the linear stability analysis are shown in Fig.\ref{third}(b). One can see that the stability of the third-order SDSs depends on both $w_m$ and $\beta$. Comparing with the fundamental and second-order SDSs, the stable region for third-order are small.  The numerical simulations with noise-added incident profiles are shown in Figs.\ref{third}(c) - (f).  In Figs.\ref{third}(c) and (e), the propagation are stable over 1000 times diffraction length, as predicated by the stability analysis[see Fig.\ref{third}(b)]. In Figs.\ref{third}(d) and (f) which are predicated unstable, the waves begin to oscillate over 100 times diffraction length..

Figure \ref{forth} shows results for the forth-order SDSs with different nonlocality degrees. There are four valleys in the forth-order SDSs as shown in Fig.\ref{forth}(a).  As the nonlocality degree increases, the shoulder of the forth-order SDSs increases and moves away from the interface.  The width of three valleys increases together.  The results of the linear stability analysis are shown in Fig.\ref{forth}(b). One can see that the stability of the forth-order SDSs depends on both $w_m$ and $\beta$. Comparing with the fundamental and second-order SDSs, the stable region for forth-order are much small. {\bf We have noticed that there are some noise in the curves in Fig.\ref{forth}(b), the correctness of this results are still under examination.}  The numerical simulations with noise-added incident profiles are shown in Figs.\ref{forth}(c) - (f).  In Figs.\ref{forth}(c) and (e), the propagation are stable over 1000 times diffraction length. In Figs.\ref{forth}(d) and (f) which are predicated unstable, the waves begin to oscillate over 100 times diffraction length..

In summary,we introduced surface image dark solitons supported by the interface between self-defocusing thermal media and linear media. Such solitons integrate the unique features
that are typical for surface waves supported by nonlocal nonlinear interfaces with those exhibited by dark solitons existing in nonlocal nonlinear media. Results motivate
further study to nonlocal dark solitons and surface solitons.

This research was supported by the National Natural Science
Foundation of China (Grant Nos. 10804033 and 10674050), the Program
for Innovative Research Team of Higher Education in Guangdong (Grant
No.06CXTD005), and the Specialized Research Fund for the Doctoral
Program of Higher Education (Grant No.200805740002).


\begin{thebibliography}{99}


\bibitem{Tomlinson-ol-1980}W. J. Tomlinson, "Surface wave at a nonlinear interface",
 Opt. Lett. {\bf 5}, 323-325 (1980).

\bibitem{Mihalachea-pio-1989}D. Mihalachea, M. Bertolottib and C. Sibiliab,
 "Nonlinear Wave Propagation in Planar Structures",
Prog. Opt.  {\bf 27}, 227-313(1989).


\bibitem{Makris-ol-2005}K. G. Makris, S. Suntsov, D. N. Christodoulides, G. I. Stegeman, A. Hache,
"Discrete surface solitons", Opt. Lett. {\bf 30}, 2466-2468 (2005).


\bibitem{Katashov-prl-2006}Y. V. Kartashov, V. A. Vysloukh, and L. Torner,
"Surface Gap Solitons", Phys. Rev. Lett. {\bf 96}, 073901 (2006).


\bibitem{Quirino-pra-1995}G. S. Garcia Quirino, J. J. Sanchez-Mondragon, and S. Stepanov , "Nonlinear surface optical waves in photorefractive crystals with a diffusion mechanism of nonlinearity"
Phys. Rev. A {\bf 51}, 1571-1577 (1995).

\bibitem{Cronin-Golomb-ol-1995}M. Cronin-Golomb,  "Photorefractive surface waves", Opt. Lett. {\bf 20}, 2075-2077 (1995).


\bibitem{Lazarides-pre-2008}N. Lazarides, G. P. Tsironis, and Y. S. Kivshar, "Surface breathers in discrete magnetic metamaterials",
Phys. Rev. E {\bf 77}, 065601(R)(2008).

\bibitem{Skinner1991-josab}S. R. Skinner and D. R. Andersen, J. Opt. Soc. Am {\bf B8}, 759-764(1991).

\bibitem{Chen1992-pra}Yijiang Chen, "Bright and dark surface waves at a nonlinear interface", Phys. Rev. {\bf A45}, 4974-4978(1992).


\bibitem{Alfassi-2007-PRL}
B. Alfassi, C. Rotschild, O. Manela, M. Segev, and D. N.
Christodoulides, "Nonlocal Surface-Wave Solitons", Phys. Rev. Lett.
{\bf98}, 213901 (2007).


\bibitem{Alfassi-2009-PRA}
B. Alfassi, C. Rotschild, and M. Segev, "Incoherent surface solitons
in effectively instantaneous nonlocal nonlinear media", Phys. Rev. A
{\bf80}, 041808 (2009).

\bibitem{Ye-2008-PRA}
F. Ye, Y. V. Kartashov, and L. Torner, "Nonlocal surface dipoles and
vortices", Phys. Rev. A {\bf77}, 033829 (2008).

\bibitem{V.Kartashov-2009-OL}
Y. V. Kartashov, V. A. Vysloukh, and L. Torner, "Multipole surface
solitons in thermal media", Opt. Lett. {\bf34}, 283-285(2009).


	
\bibitem{Kartashov2007-ol}Y. V. Kartashov, F. Ye, V. A. Vysloukh, and L. Torner, "Surface waves in defocuding thermal media," Opt. Lett. {\bf 32}, 2260-2262(2007)

\bibitem{Kartashov2007-OE}
Y. V. Kartashov, V. A. Vysloukh, and L. Torner, ``Ring surface waves
in thermal nonlinear media, Opt. Express {\bf 15,} 16216-16221 (2007),


\bibitem{Nikolov2004-OL} N. I. Nikolov, D. Neshev, W. Krolikowski, O. Bang, J. J. Rasmussen,
and P. L. Christiansen, "Attraction of nonlocal dark optical solitons,'' Opt. Lett.  {\bf 29,} 286-288 (2004).


\bibitem{Yaroslav2007-ol} Yaroslav V. Kartashov and Lluis Torner, "Gray spatial solitons in nonlocal nonlinear media," Opt. Lett. {\bf 32}, 946-948 (2007)

\bibitem{Ouyang2009-oe} S.Ouyang and Q.Guo, "Dark and gray spatial optical solitons in
Kerr-type nonlocal media," Opt. Express {\bf 17}, 5170-5175(2009)

\bibitem{Dreischuh2006-PRL}
A. Dreischuh, D. N. Neshev, D. E. Petersen, O. Bang, and W.
Krolikowski, "Observation of attraction between dark solitons,''
\prl {\bf 96,} 043901 (2006).

\bibitem{Fischer2006-ol} R. Fischer, D. N. Neshev, W. Krolikowski, Y. S. Kivshar, D. Iturbe-Castillo, S. Chavez-Cerda, M. R. Meneghetti, D. P. Caetano, and J. M. Hickman, "Oblique interaction of spatial dark-soliton stripes in nonlocal media," Opt. Lett. {\bf 31}, 3010-3012 (2006)

\bibitem{Ghofraniha2007-PRL}
N. Ghofraniha, C. Conti, G. Ruocco and S. Trillo, ``Shocks in
nonlocal media,'' Phys. Rev. Lett. {\bf 99,} 043903 (2007).

\bibitem{Conti2009-PRL}
C. Conti, A. Fratalocchi, M. Peccianti, G. Ruocco, and S. Trillo,
``Observation of a gradient catastrophe generating solitons,'' Phys. Rev. Lett.
{\bf 102,} 083902 (2009).

\bibitem{Yang2011-OE}Z. Yang, X. Ma, D. Lu, Y. Zheng, X. Gao, and W. Hu, "Relation between surface solitons and bulk solitons in nonlocal nonlinear media," Opt. Express {\bf 19}, 4890-4901 (2011).

\bibitem{Armaroli2009-pra}A. Armaroli, S. Trillo, and A. Fratalocchi, "Suppression of transverse instabilities of dark solitons and their dispersive shock waves," Phys. Rev. A {\bf 80}, 053803(2009).






\end{thebibliography}
\end{document}